\documentclass[conference,a4paper]{IEEEtran}

\usepackage[T1]{fontenc}
\usepackage[utf8]{inputenc}
\usepackage[english]{babel}
\usepackage{blindtext}
\usepackage[caption=false]{subfig}
\usepackage{algorithm}
\usepackage{algpseudocode}
\algtext*{EndFor}
\algtext*{EndWhile}
\algtext*{EndIf}
\usepackage{multirow}
\usepackage{boldline}
\usepackage{listings}
\usepackage{tcolorbox}
\usepackage{amsmath} %
\usepackage{booktabs}

\newcommand{\TODO}[1]{}

\newcommand{\mypdffig}[2]{
	\begin{figure}
		\centering
		\includegraphics[width=0.47\textwidth]{figures/#1.pdf}
		\vspace{-0.2cm}
		\caption{#2}
		\vspace{-0.15cm}
		\label{fig:#1}
	\end{figure}
}

\begin{document}

\title{Prototyping QoE‑Aware Rate Adaptation in Cellular Networks with Commercial Applications}

\author{
	\IEEEauthorblockN{
		Szilveszter N\'adas\IEEEauthorrefmark{1},
		Lars Ernstr\"om\IEEEauthorrefmark{1},
		Dan Druta\IEEEauthorrefmark{3},
		Igor Pruzhansky\IEEEauthorrefmark{4},
		David Lindero\IEEEauthorrefmark{5},\\
		Jonathan Lynam\IEEEauthorrefmark{1},
		Eric Petajan\IEEEauthorrefmark{4}
	}\
	\IEEEauthorblockA{
		\IEEEauthorrefmark{1}Ericsson Research, Santa Clara, California, USA;
		\IEEEauthorrefmark{3}AT\&T, Bothell, WA, USA;\\
		\IEEEauthorrefmark{4}AT\&T, New York, USA;
		\IEEEauthorrefmark{5}Ericsson Research, Lule{\aa}, Sweden\\
		Email: szilveszter.nadas@ericsson.com, ep619a@att.com
	}\\[-5.8ex]
}

\maketitle

\begin{figure}[!b]
\noindent\footnotesize\raggedright
\copyright~2026 IEEE. Personal use of this material is permitted. Permission from IEEE must be obtained for all other uses, in any current or future media, including reprinting/republishing this material for advertising or promotional purposes, creating new collective works, for resale or redistribution to servers or lists, or reuse of any copyrighted component of this work in other works.
\par\smallskip
Accepted author manuscript. Published in: \emph{2026 18th International Conference on Quality of Multimedia Experience (QoMEX)}, Cardiff, Wales, UK, June 2026, pp.~1--7. DOI: 10.1109/QoMEX69967.2026.11618317
\end{figure}

\begin{abstract}
Prior work has shown that QoE-aware resource sharing for real-time interactive video can support up to three times more simultaneous sessions at acceptable quality compared to rate-fair allocation.
However, the required capabilities (QoE-targeted encoding, runtime spatial complexity estimation, and rich application--network APIs) are not yet available in commercial deployments.
In this paper, we take an evolutionary approach: we design a system that delivers QoE-aware resource allocation using only capabilities that can be assembled in a lab today.
We extend the utility-based allocation framework to the radio resource domain by introducing composite spatial complexity, which combines a session's video spatial complexity with its time-variant spectral efficiency into a single resource demand function.
To operate with commercial real-time video streaming applications that use rate-based congestion control and lack capability to measure QoE, we use external tooling for QoE measurements.
We develop an incremental reallocation algorithm with per-interval limits that encode both the congestion control algorithm's speed constraint and that spatial complexity estimates are reliable only near the current rate.
The resulting prototype combines external QoE measurements with congestion-signal-based rate steering and does not require modification to commercial applications.
We chart an evolution path from this prototype toward full QoE-aware resource sharing, mapping emerging standards (IETF SCONE, CAMARA, Media over QUIC) to the progressive capabilities they enable.
\end{abstract}

\section{Introduction}
\label{sec:intro}

For mobile network operators, Quality of Experience (QoE) is becoming a competitive differentiator.
Studies show that 4 in 10 mobile video streaming sessions in the US do not yet deliver excellent quality, with up to 50\% QoE variation between operators in the same market~\cite{ericsson2024qoe}.
As video continues to dominate mobile data traffic, the ability to deliver consistently high QoE directly impacts customer retention and premium positioning.
Traditional network performance metrics such as coverage and speed tests are insufficient; what matters is the \textit{combined} quality delivered jointly by application and network, as experienced by the user.
Maximizing this experienced quality requires collaboration between application providers and network operators, precisely the application--network cooperation that the QoE-aware system proposed in this paper is designed to enable.
A recent white paper from the Video Quality Experts Group (VQEG)~\cite{vqeg2026whitepaper} presents an industry consensus on this direction, proposing a shared metric framework for structured QoE information exchange between application and network providers.

Real-time interactive (RT+I) video streaming is emerging as a key traffic category for cellular networks.
Use cases such as cloud gaming, Extended Reality (XR), teleoperation of vehicles and drones, and video conferencing all demand simultaneously high throughput and low end-to-end latency.
Unlike on-demand streaming services based on DASH/HLS, which can absorb network variations through deep client-side buffering, RT+I applications transmit each video frame as soon as it is encoded and tolerate only tens of milliseconds of delay, leaving very little room for buffering or retransmission.
The importance of both downlink and uplink video is growing: cloud gaming and XR stream video to the user, while teleoperation, teleconferencing, and emerging augmented reality glasses stream video from the user, making uplink capacity an increasingly critical resource for current 5G and future 6G networks.

The Quality of Experience of a video application is modeled using three components~\cite{nadas2024toqoe}: \textit{spatial} quality, measuring picture fidelity as affected by encoding compression; \textit{temporal} quality, capturing the fluidity of the video and degradation from stalling or frame drops; and \textit{input} quality, measuring the responsiveness of the system to user actions.
The relationship between spatial QoE and the required encoding bitrate is governed by the \textit{Spatial Complexity Curve (SCC)}, which maps bitrate to a spatial quality indicator such as Netflix's VMAF~\cite{li2016vmaf}.
This relationship exhibits a diminishing-return characteristic: beyond a certain bitrate, additional throughput yields little perceptible improvement.
Critically, the SCC varies significantly across different video content~\cite{nadas2024toqoe} and also over time within a single session as the content changes~\cite{nadas2025qoeaware}.

These properties motivate a shift from traditional bitrate-fair traffic management to \textit{QoE-aware resource allocation}.
In~\cite{nadas2024toqoe}, it was shown that exploiting the large differences in spatial complexity across sessions can yield up to $3\times$ more simultaneous streams at acceptable quality.
In~\cite{nadas2025qoeaware}, a utility-based resource sharing framework was introduced that dynamically adapts to time-variant spatial complexity, demonstrating gains over both static rate allocation and equal-QoE strategies.
The ultimate vision is a cellular radio network where resource control is optimized for user utility: a QoE controller determines a target QoE for each video stream based on information about its spatial complexity and the spectral efficiency of its radio channel.

Realizing this vision in full requires capabilities that are not yet widely available: QoE-targeted encoding, two-way application--network APIs with frequent (${\sim}1$\,s) interaction, runtime estimation of the SCC, and cooperative, trustworthy applications.
Commercial video streaming applications today do not support direct QoE targeting; they rely on well-established congestion control algorithms (CCAs) and adaptation mechanisms that operate in the bitrate domain.
Nevertheless, the network can \textit{influence} a commercial application's rate through congestion signals.
Scalable congestion control~\cite{rfc9331} and Active Rate Management~\cite{koen2025arm} within a split-responsibility scheduler~\cite{antonioli2020splitRS} provide the means to steer each session toward a target bitrate without requiring application modification.
Standardization efforts such as IETF SCONE (Standard Communication with Network Elements)~\cite{ietf-scone-protocol-04} are also defining on-path mechanisms for the network to communicate throughput advice to applications.

In this paper, we propose an evolutionary approach to QoE-aware rate adaptation that works with commercial applications available today.
Using AMVOTS (Automated Mobile Video Objective Testing System)~\cite{petajan2025amvots,banek2026kqi} as a prototyping QoE source, we obtain per-session spatial QoE indicators and bitrate measurements that are fed to a QoE controller.
The controller combines these with radio-side measurements of spectral efficiency and resource utilization to compute per-session rate targets.
Because the full SCC is not available at runtime and commercial CCAs only permit incremental rate changes, we develop an incremental reallocation algorithm that operates within these practical constraints.

The remainder of the paper is organized as follows.
Section~II presents the ideal QoE-aware resource allocation algorithm extended to the radio resource domain with spectral efficiency.
Section~III develops the incremental reallocation algorithm designed for prototyping with commercial applications.
Section~IV discusses open challenges: spatial complexity estimation, privacy, and trust.
Section~V looks beyond the prototype toward production deployment, covering evolution stages, network impairments, and enabling standards, and Section~VI concludes.

\section{QoE-aware resource allocation with spectral efficiency}
\label{sec:se-ideal}

In \cite{nadas2025qoeaware}, QoE-aware resource sharing is demonstrated for a fixed capacity in Mbps.
We extend this to the radio resource domain by incorporating time-variant spectral efficiency into the allocation algorithm.

Figure~\ref{fig:arch-ideal} shows an ideal system architecture with rich APIs. 
The QoE controller receives spatial complexity and spectral efficiency information for each session frequently. 
It also gets information about the total resources available for RT+I traffic.
Based on this information, it determines the per-session QoE target and communicates that to the streamer. 
The streamer encodes the video stream accordingly using a QoE-targeted encoder, and it sends each video frame over the transport.

\mypdffig{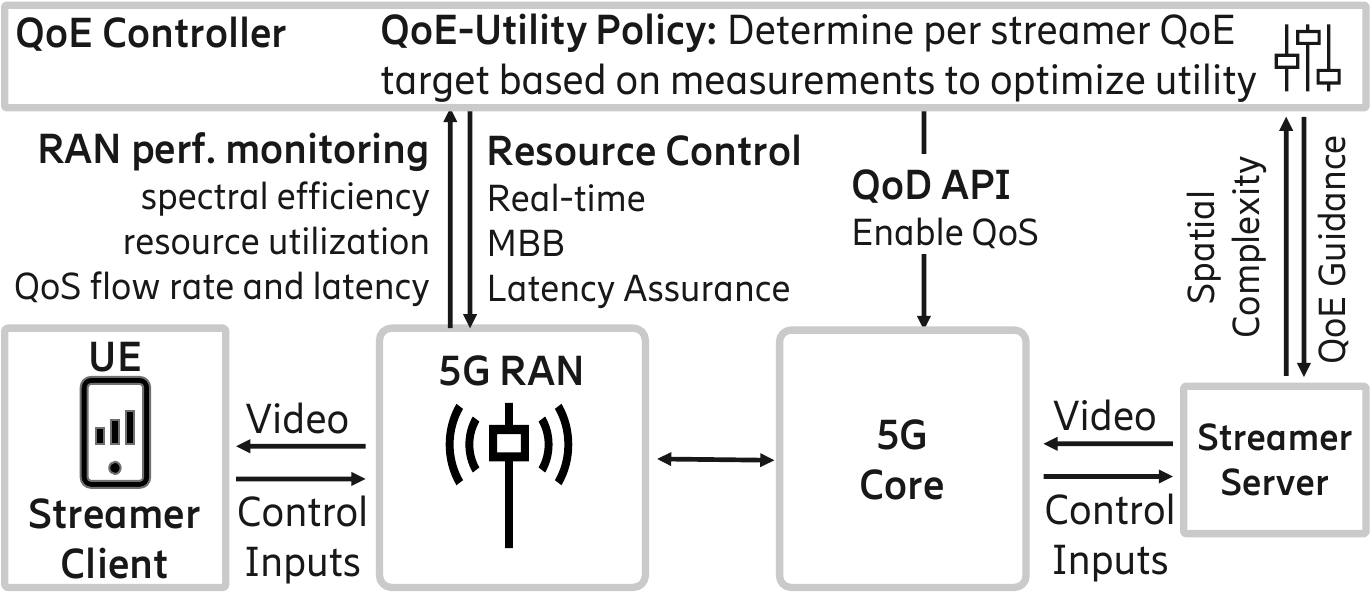}{APIs in the ideal case}
\mypdffig{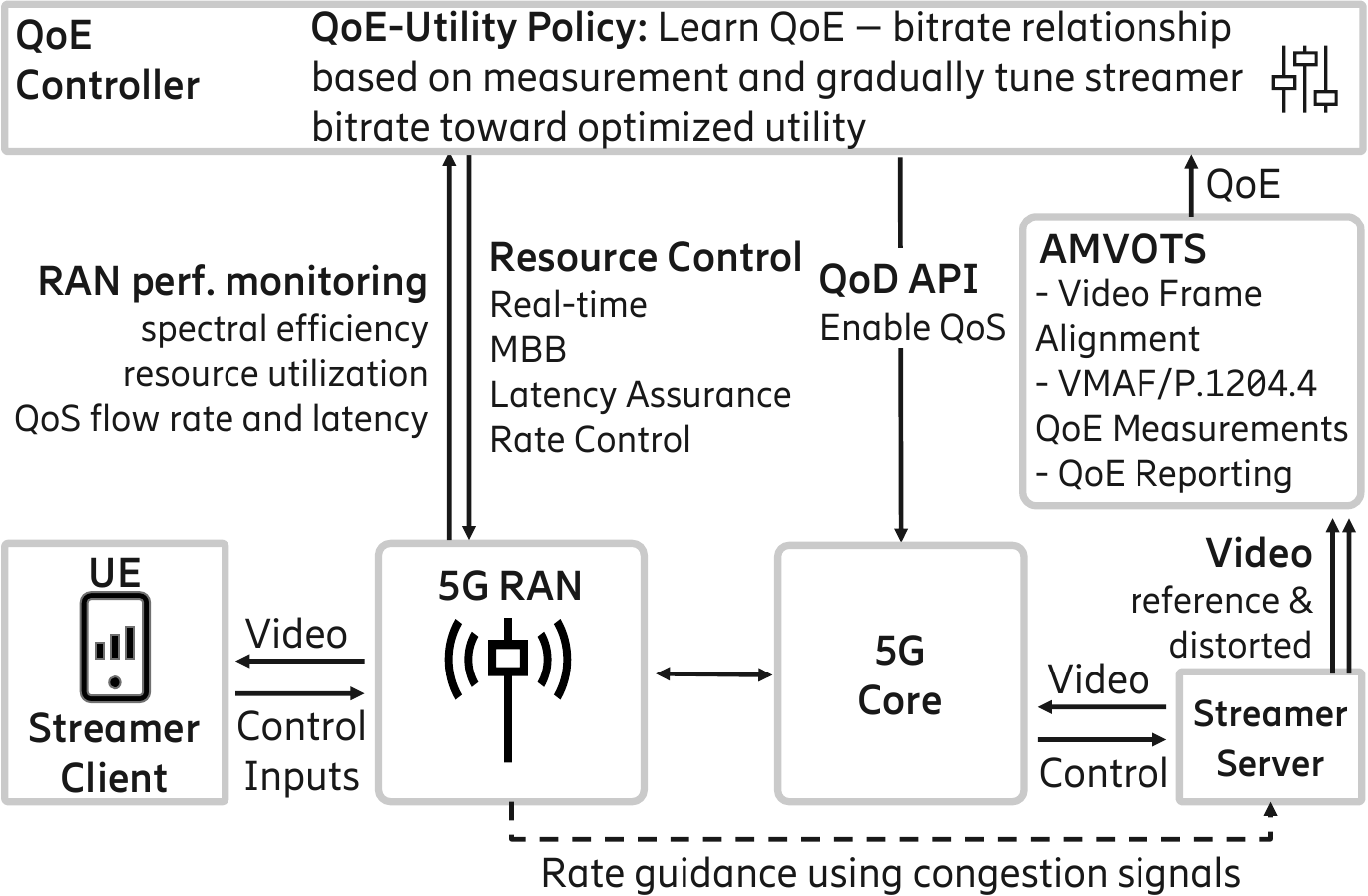}{Prototype architecture}

\newtcolorbox{algorithmbox}{
	colback=gray!10,
	colframe=black,
	sharp corners,
	boxrule=0.5pt,
	left=0pt,
	right=0pt,
	top=1pt,
	bottom=0pt,
	boxsep=1pt,
	before skip=0pt,
	after skip=0pt,
}

\begin{table}[h]
	\centering
	\begin{tabular}{lp{6cm}}
		\textbf{Symbol} & \textbf{Description} \\
		\hline
		$\Phi_{RT}(t)$ & Total radio resources for RT traffic at time $t$\\
		$\phi$ & Remaining (unallocated) radio resources \\
		$s$, $\mathcal{S}$ & A video session $s$ and the set $\mathcal{S}$ of all sessions \\
		$U(v)$ & Utility function at VMAF $v$ \\
		$V_{\min}, V_{\max}$ & Minimum and maximum considered VMAF \\
		$s.v$ & Current VMAF target for session $s$ \\
		$s.v_{\text{next}}$ & Next candidate VMAF for session $s$ \\
		$s.\phi$ & Current allocated radio resources for $s$ \\
		$s.\eta(t)$ & Spectral efficiency for $s$ at time $t$ \\
		$s.R(t, v)$ & Rate required for $s$ at time $t$ to reach VMAF=$v$ (spatial complexity) \\
		$s.\Phi(t,v)$ & Radio resources required for $s$ at time $t$ to reach VMAF=$v$ (composite spatial complexity)\\
		$s.c^+$ & Whether $s.v$ (and thus $s.\phi$) can still be increased \\
		$s.\delta u$ & Marginal utility for $s$ in the radio resource domain \\
	\end{tabular}
	\vspace{0.2cm}
	\caption{Notations used in Algorithm~\ref{alg:max-util-se}.}
	\vspace{-0.5cm}
	\label{t:notations_se}
\end{table}

For a session $s \in \mathcal{S}$, the spatial complexity $s.R(t,v)$ describes the rate required to achieve a given VMAF, which varies with video content over time.
When considering radio networks, the spectral efficiency $s.\eta(t)$ determines how much rate can be achieved per unit of radio resource (e.g., a physical resource block in 5G NR).
We define the \textit{composite spatial complexity} as:
\begin{equation}
	s.\Phi(t,v) = \frac{s.R(t, v)}{s.\eta(t)}
\end{equation}
This combines the video's spatial complexity with the channel's spectral efficiency into a single time-variant function representing the radio resources required to achieve a given VMAF.

Algorithm~\ref{alg:max-util-se} (notation in Table~\ref{t:notations_se}) extends the maximum utility resource allocation from \cite{nadas2025qoeaware} to operate in the radio resource domain, taking spectral efficiency into account.
The algorithm distributes $\phi = \Phi_{RT}(t)$ radio resources to maximize total utility.
The key change from the rate-domain algorithm is that marginal utility $s.\delta u$ is computed per unit of radio resource rather than per unit of rate.
The utility function $U(v)$ encodes the operator's resource sharing policy.
We use a piecewise-linear concave curve with three anchors: $V_{\min}$ (admission threshold), $V_{\text{desired}}$ (target quality), and $V_{\max}$ (above which utility is flat).
Decreasing slopes between anchors give a priority ordering: admitting a session at $V_{\min}$ is worth several upgrades to $V_{\text{desired}}$, and each such upgrade is worth several luxury improvements toward $V_{\max}$.
The operator tunes the trade-off by moving the anchors; full design is detailed in~\cite{nadas2025qoeaware}.

\begin{algorithm}
	\caption{Maximum Utility Resource Allocation with Spectral Efficiency}
	\label{alg:max-util-se}
	
	\begin{algorithmbox}
		\begin{algorithmic}[1]
			\State $\phi \gets \Phi_{RT}(t)$ \Comment{Initialize remaining radio resources}
			\ForAll{$s \in \mathcal{S}$}
			\State $s.v \gets 0$
			\State $s.v_{\text{next}} \gets V_{\min}$
			\State $s.\phi \gets 0$
			\State $s.c^+ \gets \text{true}$
			\State $s.\delta u \gets \frac{U(s.v_{\text{next}})}{s.\Phi(t, s.v_{\text{next}})}$
			\EndFor
			\While{True}
			\State $s \gets \arg\max_{q \in \mathcal{S}} q.\delta u$ \Comment{Session with highest $\delta u$}
			\If{$\lnot\, s.c^+$} \textbf{break} \Comment{No session can increase}
			\EndIf
			\State $\Delta \phi \gets s.\Phi(t, s.v_{\text{next}}) - s.\phi$
			\If{$\phi < \Delta \phi$} \Comment{Not enough free resources}
			\State $s.c^+ \gets \text{false}$
			\State $s.\delta u \gets 0$
			\State \textbf{continue}
			\EndIf
			\State $\phi \gets \phi - \Delta \phi$
			\State $s.v \gets s.v_{\text{next}}$
			\State $s.\phi \gets s.\Phi(t, s.v)$
			\State $s.v_{\text{next}} \gets s.v + 1$
			\If{$s.v_{\text{next}} > V_{\max}$}
			\State $s.c^+ \gets \text{false}$
			\State $s.\delta u \gets 0$
			\Else
			\State $s.\delta u \gets \frac{U(s.v_{\text{next}}) - U(s.v)}{s.\Phi(t, s.v_{\text{next}}) - s.\Phi(t, s.v)}$
			\EndIf
			\EndWhile
		\end{algorithmic}
	\end{algorithmbox}
\end{algorithm}

The algorithm outputs the target VMAF $s.v$ for each session, communicated to the application.
It also determines the target long-term radio resources $s.\phi$ used by the session.
The corresponding session rate is $s.r = s.\phi \cdot s.\eta(t)$.

\section{Incremental QoE-aware rate adaptation}
\label{sec:incremental}

Since the capabilities to support the ideal system (Section~\ref{sec:se-ideal}) are not yet available, we prototype QoE-aware resource allocation using the lab setup shown in Figure~\ref{fig:arch-amvots}.
AMVOTS~\cite{petajan2025amvots,banek2026kqi} measures QoE per session externally, in place of in-application reporting, and the QoE controller computes per-session resource targets.
The system operates on two timescales.
The QoE controller updates per-session targets every interval (${\sim}1$\,s).
On a per-RTT basis, a split-responsibility scheduler~\cite{antonioli2020splitRS} running Active Rate Management~\cite{koen2025arm} enforces each target by generating congestion signals that steer the session's L4S~\cite{rfc9331} congestion control.

Algorithm~\ref{alg:max-util-se} assumes full knowledge of the composite SCC and can set arbitrary allocations in one step.
Two constraints prevent this in practice, and each motivates a per-interval limit.
First, only a single measured (rate, VMAF) pair is available per session per interval, yielding one point on the composite SCC, not the full curve.
The VMAF limit $\Delta V_{\max}$ (e.g., 5 points per interval) bounds how far the algorithm moves from this operating point, keeping it where the SCC estimate is reliable.
The algorithm adjusts VMAF in steps of one point per iteration; finer steps would add complexity without meaningful benefit, as perceptual differences require several VMAF points~\cite{vmaf-jnd}.
Second, since the CCA can only change rate incrementally, the resource limits $\rho^+$ and $\rho^-$ (e.g., 1.1 and 0.7) bound the ratio between new and previous allocation, reflecting that CCAs probe upward cautiously but can cut rate quickly in response to congestion signals.
The first VMAF change per interval is always permitted regardless of $\rho$, ensuring at least one adjustment per session.
With $\Delta V_{\max} = 5$, a session converges within a few intervals; the resulting slow pace of change is by design, favoring QoE stability over fast adaptation.

A session at $V_{\min}$ cannot decrease further (quality below $V_{\min}$ is unacceptable); decreasing would waste resources on a session delivering unacceptable quality.
Instead, when a session is at or below $V_{\min}$ and $\Delta V_{\max}$ allows a further step down, UpdateMarginals (Algorithm~\ref{alg:update-marginals}) offers a \textit{drop}: the session is released from the QoE-managed set and falls back to best-effort service in the mobile broadband pool, freeing all its reserved resources.
The drop is executed under the same conditions as any decrease (resource pressure or net utility gain) and bypasses $\rho^-$ because the session is released, not gradually slowed.
Dropped sessions are excluded from $\mathcal{S}$ for a cooldown period (e.g., 1\,min) to prevent oscillation; after cooldown, re-admission follows the normal path if resources permit.

Algorithm~\ref{alg:incremental} (Table~\ref{t:notations_alg2}) initializes each session from measurements and then iteratively reallocates resources.
New sessions join $\mathcal{S}$ once measured and compete for admission via the same mechanism; if resources do not permit reaching $V_{\min}$, the session remains at its CCA-determined quality and may be released under the same conditions as any other session.
The free resource budget $\phi_{\text{free}}$ can be negative if $\Phi_{RT}(t)$ decreased since the previous interval or if a session's spectral efficiency dropped, in which case the algorithm must decrease sessions to restore feasibility.
The UpdateMarginals subroutine (Algorithm~\ref{alg:update-marginals}) evaluates each session's next feasible increase and decrease step.
It tracks how far the session's VMAF has moved from its measured value this interval ($\Delta v = s.v - \text{round}(s.\tilde{v}(t-1))$); this is the quantity bounded by $\Delta V_{\max}$.
For each direction, it computes a target VMAF ($s.v_{\text{next}}$), the associated resource cost or savings ($s.\Delta\phi$), a feasibility flag ($s.c$), and the marginal utility $s.\delta u = (U(s.v_{\text{next}}) - U(s.v)) / s.\Delta\phi$.
This marginal utility gives utility per unit of radio resource, directly comparable across sessions regardless of their spectral efficiency or spatial complexity.
A step is feasible only if the $\Delta V_{\max}$ budget allows it and the resulting allocation stays within the $\rho$ bounds; blocked sessions receive $s.\delta u^+ = 0$ or $s.\delta u^- = -\infty$, naturally losing the argmax.
UpdateMarginals also handles admission (sessions below $V_{\min}$ target $V_{\min}$ directly, bypassing per-interval limits because restoring acceptable quality takes priority over CCA convergence) and the drop described above.

The main loop selects the session $s^+$ with the highest $s.\delta u^+$ and increases it if resources are available.
Otherwise, it selects $s^-$ (smallest loss) and decreases it if the resource budget is exceeded ($\phi_{\text{free}} < 0$) or if reallocating from $s^-$ to $s^+$ increases total utility ($s^+.\delta u^+ > |s^-.\delta u^-|$).
The loop terminates when no beneficial reallocation remains; the output is the target VMAF $s.v$ and resource allocation $s.\phi$ for each session.

\begin{table}[h]
	\centering
	\begin{tabular}{lp{5.5cm}}
		\textbf{Symbol} & \textbf{Description} \\
		\hline
		$\phi_{\text{free}}$ & Free (unallocated) radio resources \\
		$s.\tilde{\phi}(t-1)$ & Measured resource usage for $s$ in previous interval \\
		$s.\tilde{v}(t-1)$ & Measured VMAF for $s$ in previous interval \\
		$s.\tilde{\eta}(t-1)$ & Measured spectral efficiency for $s$ \\
		$\Delta v$ & VMAF change from measured value this interval \\
		\hline
		\multicolumn{2}{l}{\textit{Per-interval limits (inputs):}} \\
		$\Delta V_{\max}$ & Max VMAF change per interval (e.g., 5) \\
		$\rho^+$ & Max resource ratio after increase (e.g., 1.1) \\
		$\rho^-$ & Min resource ratio after decrease (e.g., 0.7) \\
		\hline
		\multicolumn{2}{l}{\textit{Per-session outputs of UpdateMarginals:}} \\
		$s.c^+$, $s.c^-$ & Whether increase / decrease of $s.v$ is feasible \\
		$s.v_{\text{next}}^+$, $s.v_{\text{next}}^-$ & VMAF after increase / decrease \\
		$s.\delta u^+$, $s.\delta u^-$ & Marginal utility for increase / decrease \\
		$s.\Delta\phi^+$, $s.\Delta\phi^-$ & Resources needed / freed \\
	\end{tabular}
	\vspace{0.2cm}
	\caption{Additional notations for Algorithms~\ref{alg:incremental} and~\ref{alg:update-marginals} (see also Table~\ref{t:notations_se}).}
	\vspace{-0.5cm}
	\label{t:notations_alg2}
\end{table}

\begin{algorithm}
	\caption{Incremental Utility-Based Resource Reallocation}
	\label{alg:incremental}
	
	\begin{algorithmbox}
		\begin{algorithmic}[1]
			\ForAll{$s \in \mathcal{S}$} \Comment{Initialize from measurements}
			\State $s.\phi \gets s.\tilde{\phi}(t-1)$
			\State $s.v \gets \text{round}(s.\tilde{v}(t-1))$
			\State \Call{UpdateMarginals}{$s$}
			\EndFor
			\State $\phi_{\text{free}} \gets \Phi_{RT}(t) - \sum_{s \in \mathcal{S}} s.\phi$ \Comment{Can be negative}
			\While{True} \Comment{Main loop}
			\State $s^+ \gets \arg\max_{q \in \mathcal{S}} q.\delta u^+$ \Comment{Largest gain}
			\If{$s^+.c^+ \land \phi_{\text{free}} \geq s^+.\Delta\phi^+$} \Comment{Increase}
			\State $\phi_{\text{free}} \gets \phi_{\text{free}} - s^+.\Delta\phi^+$
			\State $s^+.\phi \gets s^+.\phi + s^+.\Delta\phi^+$
			\State $s^+.v \gets s^+.v_{\text{next}}^+$
			\State \Call{UpdateMarginals}{$s^+$}
			\State \textbf{continue}
			\EndIf
			\State $s^- \gets \arg\max_{q \in \mathcal{S}} q.\delta u^-$ \Comment{Smallest loss}
			\Statex \Comment{Over budget or beneficial swap}
			\State $\text{shouldDec} \gets \phi_{\text{free}} < 0 \lor s^+.\delta u^+ > |s^-.\delta u^-|$
			\If{$s^-.c^- \land \text{shouldDec}$} \Comment{Decrease}
			\State $\phi_{\text{free}} \gets \phi_{\text{free}} + s^-.\Delta\phi^-$
			\State $s^-.\phi \gets s^-.\phi - s^-.\Delta\phi^-$
			\State $s^-.v \gets s^-.v_{\text{next}}^-$
			\State \Call{UpdateMarginals}{$s^-$}
			\State \textbf{continue}
			\EndIf
			\If{$\lnot\, s^+.c^+$} \textbf{break} \Comment{No session can increase}
			\EndIf
			\State $s^+.c^+ \gets \text{false}$ \Comment{Mark $s^+$: cannot increase}
			\State $s^+.\delta u^+ \gets 0$
			\EndWhile
		\end{algorithmic}
	\end{algorithmbox}
\end{algorithm}

\begin{algorithm}
	\caption{UpdateMarginals($s$)}
	\label{alg:update-marginals}

	\begin{algorithmbox}
		\begin{algorithmic}[1]
			\State $\Delta v \gets s.v - \text{round}(s.\tilde{v}(t-1))$ \Comment{VMAF change this interval}
			\Statex \Comment{Increase}
			\If{$0 < s.v < V_{\min}$} \Comment{Admission: target $V_{\min}$}
			\State $s.v_{\text{next}}^+ \gets V_{\min}$
			\State $s.\Delta\phi^+ \gets s.\Phi(t, V_{\min}) - s.\phi$
			\State $s.c^+ \gets s.\Delta\phi^+ > 0$
			\ElsIf{$s.v < V_{\max} \land |\Delta v + 1| \leq \Delta V_{\max}$} \Comment{Normal}
			\State $s.v_{\text{next}}^+ \gets s.v + 1$
			\State $s.\Delta\phi^+ \gets s.\Phi(t, s.v+1) - s.\phi$
			\State $s.c^+ \gets \Delta v = 0 \lor s.\phi + s.\Delta\phi^+ \leq \rho^+ \cdot s.\tilde{\phi}(t-1)$ \Comment{First step free, then $\rho$-bounded}
			\Else \Comment{Blocked}
			\State $s.c^+ \gets \text{false}$
			\State $s.\delta u^+ \gets 0$
			\EndIf
			\If{$s.c^+$}
			\State $s.\delta u^+ \gets \frac{U(s.v_{\text{next}}^+) - U(s.v)}{s.\Delta\phi^+}$
			\EndIf
			\Statex \Comment{Decrease}
			\If{$s.v > V_{\min} \land |\Delta v - 1| \leq \Delta V_{\max}$} \Comment{Normal}
			\State $s.v_{\text{next}}^- \gets s.v - 1$
			\State $s.\Delta\phi^- \gets s.\phi - s.\Phi(t, s.v-1)$
			\State $s.c^- \gets \Delta v = 0 \lor s.\phi - s.\Delta\phi^- \geq \rho^- \cdot s.\tilde{\phi}(t-1)$ \Comment{First step free, then $\rho$-bounded}
			\ElsIf{$0 < s.v \leq V_{\min} \land |\Delta v - 1| \leq \Delta V_{\max}$}
			\State $s.v_{\text{next}}^- \gets 0$ \Comment{Drop}
			\State $s.\Delta\phi^- \gets s.\phi$ \Comment{Free all resources}
			\State $s.c^- \gets s.\phi > 0$
			\Else \Comment{Blocked}
			\State $s.c^- \gets \text{false}$
			\State $s.\delta u^- \gets -\infty$
			\EndIf
			\If{$s.c^-$}
			\State $s.\delta u^- \gets \frac{U(s.v_{\text{next}}^-) - U(s.v)}{s.\Delta\phi^-}$ \Comment{Negative}
			\EndIf
		\end{algorithmic}
	\end{algorithmbox}
\end{algorithm}

\section{Discussion}

\subsection{Spatial complexity estimation}

Both algorithms use the spatial complexity curve $s.R(t,v)$ as input.
In the ideal architecture (Figure~\ref{fig:arch-ideal}), the streamer determines the SCC directly: the encoder can explore encoding with different settings, and intermediate data computed during encoding can be used for SCC estimation.
In the prototype (Figure~\ref{fig:arch-amvots}), neither the streamer nor the network has access to the full curve.
AMVOTS provides a single measured (rate, VMAF) pair per session per interval; combined with the measured spectral efficiency, this yields one point on the composite SCC, not the full curve, and not even its local slope.
Algorithm~\ref{alg:incremental} requires the SCC within $\pm 5$ VMAF values ($\Delta V_{\max}$) of the current operating point to compute resource costs, so the local slope must be estimated.
As the algorithm adjusts sessions across intervals, each session's operating point moves, tracing out nearby points on the SCC that can be used to build a local estimate (e.g., linear extrapolation from the two most recent operating points); the exact estimation algorithm is left for future work.
The per-interval limits $\Delta V_{\max}$ and $\rho$ confine each step to a small neighborhood around the current working point, keeping this estimation problem tractable.

Even in a fully evolved system with rich SCC information, limiting the pace of QoE change remains desirable: it keeps QoE stable for the user and reduces the SCC knowledge required to a small neighborhood.
Developing the actual SCC estimation algorithm (which may combine model-based and data-driven methods) is an important open challenge but outside the scope of this paper.

\subsection{Privacy}

For RT+I video, each session typically produces unique content, making it difficult to identify what is being streamed from the QoE reporting alone.
This assumption breaks when the stream contains recorded content with a known QoE fingerprint, leading to concerns regarding user privacy.
In such cases, noise can be injected by occasionally over-representing the spatial complexity, for example by inflating QoE reports or increasing the encoder's quality target.
Because the noise always represents higher complexity, quality targets can still be met; the cost is a small reduction in total utility.

\subsection{Trust and incentives}

QoE-aware resource allocation creates an incentive for applications to misrepresent their QoE, spatial complexity, or compliance with guidance to obtain more resources.
Creating the right economic incentives to discourage this is a hard problem.
A complementary technical approach is for network vendors to provide a trusted SDK that handles video encoding, transport, and API communication, ensuring that reported metrics are reliable and guidance is followed.

\section{Beyond the Prototype}

Bridging the gap between the prototype (Figure~\ref{fig:arch-amvots}) and the ideal system (Figure~\ref{fig:arch-ideal}) requires progressively richer capabilities, each addressing a broader set of impairments and enabled by ongoing standardization.

\subsection{Evolution stages}

In the \textit{prototype} stage, the QoE controller relies on external measurements from AMVOTS and steers sessions through congestion signals.
Because the QoE controller treats per-session QoE as a typed input, the prototype generalizes across RT+I content (cloud gaming, XR, teleoperation) once the corresponding application supplies in-application QoE measurements.
As the first step beyond the prototype, QoE measurements need to move into the streaming application itself, enabling production deployment without external measurement infrastructure.

\textit{Rate guidance} overcomes the limitations of rate-based CCAs, which may not ramp up fast enough when spatial complexity increases.
Explicit rate guidance messages from the network enable fast and precise rate adjustment, removing the $\rho$ constraint from Algorithm~\ref{alg:incremental}.

\textit{QoE guidance} goes one step further: the streamer accepts QoE targets instead of rate targets.
Controlling QoE directly, rather than indirectly through rate, makes the control loop more stable.
However, this trades burstiness in QoE for burstiness in rate and resources, which current networks are not designed to handle~\cite{nadas2025qoeaware}.
When a streamer accepts QoE guidance, it must also implement it in the encoder.
The ideal solution is a QoE-targeted encoder, but no such encoder is available today; it can be approximated by an internal control loop that adjusts encoder parameters to reach the target QoE.

Even in a fully evolved system, the incremental approach of Algorithm~\ref{alg:incremental} retains value.
Limiting the pace of QoE change benefits session quality, and confining changes to a small neighborhood around the current working point means the full SCC is never needed, only its local slope.

\subsection{Network impairments}

Subscriber video QoE is affected by impairments across the radio access network (RAN), core/transport, and content delivery path.
In the RAN, poor signal quality, weak coverage, cell congestion, and mobility events such as handovers reduce achievable throughput and cause resolution drops, stalling, and quality oscillations.
User equipment wake-up delays from low-activity states
increase startup time and cause quality transients during bursty traffic.
Beyond the RAN, core network latency, jitter, poor peering, and content delivery network cache misses further degrade startup time and sustained quality.
At the application layer, adaptive bitrate misprediction under rapid network fluctuations leads to overshoot or undershoot, causing stalls and quality oscillations, especially in low-latency streaming with short buffers.

These impairments operate on different timescales, and at different layers, and no single mechanism can address all of them.
The evolution stages progressively expand the QoE controller's ability to identify, attribute, and mitigate them:
\begin{itemize}
\item The \textit{prototype} stage already addresses congestion-driven QoE degradation by reallocating resources across sessions based on measured QoE and spectral efficiency.
Sessions with severely degraded channels are naturally deprioritized, freeing resources for sessions that can use them effectively.
\item \textit{Rate guidance} enables faster reaction to signal quality changes and handover events, overcoming CCA ramp-up limitations that cause prolonged quality drops after impairments.
\item \textit{QoE guidance} and \textit{QoE-targeted encoding} allow the system to handle content-driven complexity changes without rate overshoot, reducing the ABR misprediction problem at its source.
\item \textit{Application-reported QoE} (via CAMARA or in-app measurement) enables root-cause attribution: correlating QoE degradation with specific radio or transport impairments lets operators prioritize targeted optimization over broad capacity upgrades.
\item \textit{Real-time and prediction capabilities} will need to evolve to address network impairments and ensure consistent quality.
\end{itemize}
Mapping observed QoE degradation back to its root cause is essential for efficient network optimization, moving operators from simply observing poor experience to identifying and addressing the specific impairments that matter most.

\subsection{Standards and protocols enabling the evolution}

The evolutionary stages require progressively richer signaling between applications and the network.
Active standardization efforts are creating the building blocks for each stage.

IETF is finalizing SCONE~\cite{ietf-scone-protocol-04}, an on-path mechanism defined as a QUIC extension that allows applications to request throughput advice from the access network.
SCONE directly enables the \textit{rate guidance} stage: the QoE controller's per-session rate target can be delivered as SCONE throughput advice, replacing or supplementing CCA-based congestion signaling.
Because the SCONE proxy endpoint is a hop on the QUIC data path, it inherently identifies the flow and can establish the control connection to the QoE controller, eliminating the out-of-band flow-to-metadata mapping that control-plane approaches require.

Media over QUIC (MoQ)~\cite{ietf-moq-transport-17}, under development in IETF, is a publish-subscribe protocol designed for real-time media delivery over QUIC.
MoQ brings codec flexibility, simpler client-server architectures, and native support for rate adaptation through codec parameter changes.
Because MoQ is built on QUIC, SCONE throughput advice will be available to MoQ-based applications, providing a natural integration point for network-guided rate adaptation.

For the reverse direction, application-to-network QoE reporting, the Linux Foundation CAMARA initiative~\cite{camara-project} defines APIs that can be extended to expose application QoE metrics, either per-device or aggregated at cell or sector level.
This enables application-reported QoE without requiring on-path protocol changes.
Aggregated QoE data from cohorts of devices can also be correlated with network measurements to improve SCC estimation and RAN scheduling.

Within the Open RAN ecosystem, standardized interfaces for publishing network performance measurements are being developed to enable their consumption by rApps and xApps. This creates an opportunity to implement the QoE controller as an rApp/xApp, leveraging O-RAN RAN Intelligent Controller (RIC) interfaces to monitor spectral efficiency and radio resource utilization required by Algorithms 1 and 2. Operator-specific objectives, including the utility function and other QoE controller configurations, can be conveyed through an A1 policy.

As the ecosystem matures toward \textit{QoE guidance} and \textit{QoE-targeted encoding}, richer two-way APIs will be needed, combining SCONE's network-to-app throughput advice with CAMARA's app-to-network QoE reporting into a closed-loop interaction.
The trusted SDK approach described in Section~IV can bundle these protocol interactions into a single integration point for application developers, lowering the adoption barrier for the most advanced stages of the evolution.

\section{Conclusion}
\label{sec:conclusion}

QoE-aware resource allocation promises large efficiency gains for cellular networks carrying real-time interactive video, but its practical realization has remained an open problem.
The core difficulty is not algorithmic: given full knowledge of spatial complexity and direct control over encoding, a greedy utility-maximizing allocation is straightforward.
The difficulty is that commercial applications expose neither.
They use rate-based congestion control that permits only incremental rate changes, and they provide no QoE or spatial complexity information to the network.
Even with external measurement, only a single operating point on the spatial complexity curve is available per interval.

The incremental reallocation algorithm developed in this paper is designed around these two constraints.
Its per-interval limits (bounding VMAF change to the region where the SCC estimate is reliable, and bounding resource change to what the CCA can achieve) are not approximations of the ideal algorithm but a fundamentally different design that treats limited observability and limited control as first-class constraints.
The composite spatial complexity formulation extends this to the radio domain, where per-session spectral efficiency creates additional variation that rate-domain models cannot capture.

The prototype architecture demonstrates that this is realizable in a lab environment today: external QoE measurement and network-side rate steering via congestion signals achieve QoE-aware allocation without requiring any modification to the streaming application.
Moving from lab prototype to production deployment requires replacing AMVOTS with in-application QoE measurement, the first step of the evolution path.

The evolution path we outline is not speculative; the building blocks are under active standardization.
SCONE~\cite{ietf-scone-protocol-04} provides on-path rate guidance that can replace congestion-signal-based steering.
CAMARA~\cite{camara-project} enables application-to-network QoE reporting without protocol changes.
Media over QUIC offers a transport layer with native support for codec-level rate adaptation.
As these capabilities mature, the per-interval limits that Algorithm~\ref{alg:incremental} imposes can be progressively relaxed, converging toward the ideal allocation.

Key open challenges remain: runtime estimation of the spatial complexity curve from limited observations, session-level QoE models that capture the temporal effects of quality transitions, privacy-preserving mechanisms for sharing spatial complexity information, and incentive structures that discourage applications from misrepresenting their QoE to gain resource advantages.

\section*{Acknowledgments}
The authors used Anthropic's Claude models, accessed through Claude Code and Amazon Kiro, for text drafting, algorithm formalization, and notation design in Sections~\ref{sec:intro} through~\ref{sec:conclusion} of this manuscript.
The authors reviewed, verified, and edited all AI-generated content and take full responsibility for the final publication.

\bibliographystyle{IEEEtran}
\bibliography{paper}

\end{document}